# Safety Evaluation of Transit Signal Priority with Bus Speed Volatility as a Surrogate Measure: Case Study in Minnesota

[Preprint]*

Yu Song[1], Hoki Tse[2], Eric M. Lind[3], Madhav V. Chitturi[1], David A. Noyce[1]

1. University of Wisconsin-Madison, 2. IBI Group, 3. Metro Transit

**ABSTRACT**

Previous studies have found correlations between the implementation of transit signal priority (TSP) and the reduction in number of crashes. To further understand how TSP affects traffic safety, a more in-depth evaluation was carried out using detailed bus automatic vehicle location (AVL) data. The data was from Minneapolis-Saint Paul Metro Transit Bus Route 5, where TSP was implemented at 30 signalized intersections in early 2019. A surrogate safety measure, bus speed volatility (BSV), was used to estimate TSP's safety effects, with a higher BSV indicating more safety risks. A regression analysis was carried out on 23,123 event-level observations, with event defined as a bus traversal of a TSP-equipped intersection. Results indicate that with a TSP request, BSV was significantly lower than without a TSP request, confirming TSP's effectiveness in smoothing bus trips through intersections, thus reducing risks of bus collisions and passenger fall-overs.

**INTRODUCTION**

Transit signal priority (TSP) is a transit preferential treatment widely applied by transit agencies in U.S. cities. The purpose of implementing TSP is to reduce transit delay. TSP achieves that purpose by detecting transit vehicles approaching signalized intersections and adjusting traffic signal timing to provide the transit vehicles longer or early green indications. TSP implemented in U.S. cities can save bus travel time by 2% to 20%, with 8% to 12% as the most typical range (Danaher et al. 2007). TSP was found to increase side-street traffic delay, since it takes time from side-street phases to provide priority on corridor phases. Technologies such as GPS-based TSP enable flexible and conditional priority control and keep this negative impact on side-street to a minimum (Song et al. 2016a; b).

    Despite abundant research confirming the operational benefits of TSP, practitioners and researchers knew relatively little about the effects of TSP on traffic safety. There are only a handful of published studies on TSP safety (Goh et al. 2013, 2014; Li et al. 2017; Naznin et al. 2016; Shahla et al. 2009; Song and Noyce 2018, 2019). Those previous studies focused on quantifying TSP's effects on aggregated number of traffic crashes at the intersection or corridor level. Mixed results were reported, but the most recent case studies of bus TSP implementations in the U.S. found TSP effective in reducing number of crashes at the corridor level (Song and Noyce 2018, 2019).

    Prior evaluations of TSP's safety effects using aggregated crash data estimated the magnitude of TSP's effects on traffic safety. However, not enough insights were provided to help understand the mechanism of such effects. Since TSP adjusts signal timing, it is

---

reasonable to expect effects of changed signal timing on transit and traffic movement characteristics. Changes in vehicle movement characteristics, especially at intersections, is considered to be closely linked to changes in safety risks. This assumption is supported by multiple studies on crash causality and surrogate safety measures (Arvin et al. 2019; Davis 2001; Davis et al. 2011; Kamrani et al. 2018; Porter et al. 2018; Wali et al. 2018). Therefore, a hypothesis evaluated in this study is whether and how much TSP affects transit vehicle movement characteristics closely related to safety risks. A bus speed volatility (BSV) measure was defined and used as such a surrogate safety measure in this study. Bus TSP and automatic vehicle location (AVL) data from a Minneapolis-Saint Paul Metro Transit bus route were used to calculate BSV. The relationship between BSV and variables including TSP and others were evaluated using regression analysis.

**LITERATURE REVIEW**

Recent studies of Australia and U.S. TSP implementations found the technology to be effective in reducing the number of crashes at intersection and corridor levels (Goh et al. 2013, 2014; Naznin et al. 2016; Song and Noyce 2018, 2019). Goh et al. and Naznin et al. studied TSP implementations in Melbourne. A before-after study by Goh et al. showed that fatal and injury crashes, aggregated at the intersection level, reduced by 11.1% after TSP implementation (Goh et al. 2013). A cross-sectional study by Goh et al. showed that number of bus-involved crashes, aggregated at the corridor level, was lowered by 53.5% with TSP and other bus preferential treatments such as providing bus lanes and consolidating bus stops (Goh et al. 2014). A before-after study by Naznin et al. assessed streetcar TSP and found a reduction of 13.9% in number of streetcar-involved crashes at the intersection level (Naznin et al. 2016). Song and Noyce studied TSP implementations in the metropolitan areas of Seattle, WA, and Portland, OR, and found reductions in corridor-level crash numbers by 11.0% and 4.5%, respectively, after TSP implementations. (Song and Noyce 2018, 2019)

Two prior studies of Canadian implementations found that TSP may lead to a higher number of crashes. A study by Shahla et al. of bus and streetcar TSP implementations in Toronto showed that transit vehicle-involved crash numbers could be higher by as much as 40% at intersections with TSP, than at intersections without TSP (Shahla et al. 2009). Li et al. studied the same Toronto TSP implementations using VISSIM micro-simulation and Surrogate Safety Assessment Model (SSAM), and the results showed that by removing TSP from TSP-equipped intersections, the number of crashes reduced by 1.6% overall, but the change in number of transit vehicle crashes ranged from a reduction of 0.1% to an increase of 1.6% (Li et al. 2017). Li et al. concluded that implementing TSP would potentially lead to an increase in the number of crashes at intersections.

Through a review of earlier studies on TSP's safety effects, the following three research gaps were identified for this and future studies to fill:
- Earlier studies yielded mixed results about TSP's effects on safety, thus more in-depth evaluations of this topic are still needed to obtain a better understanding.
- Earlier studies only used crash numbers as the measure of safety performance. Other measures such as passenger incidents are also important safety indices used by transit agencies and can be included in TSP safety evaluations.
- Earlier studies estimated TSP's safety effects on aggregated crash numbers. Available microscopic bus operational data such as AVL would enable more detailed analysis and offer insights in the mechanism of TSP's safety effects.

This study adds to the existing safety evaluations of TSP by taking a step further in revealing the mechanism of TSP's effects on traffic safety. By applying bus speed volatility (BSV) as a surrogate safety measure, this study took both bus collision and passenger incident risks into consideration. The Minneapolis TSP evaluated in this study was implemented along with few geometric modifications and no bus stop relocated, thus controlling of confounding factors was easily done. The high-quality bus AVL data provided by Metro Transit ensured abundant samples for a valid statistical analysis.

**TSP IN THE MINNEAPOLIS AREA**

Metro Transit Bus Route 5 is a local bus route that runs north-south through several cities in the Minneapolis metropolitan area. It is one of the busiest Metro Transit bus routes and runs a high-frequency service (with a bus every 15 minutes) between North 26th Avenue and East 56th Street in the City of Minneapolis. To improve the on-time performance and reduce travel time of Route 5 buses, as well as to prepare for a future conversion of Bus Route 5 into D-Line Bus Rapid Transit (BRT) service, Metro Transit started adding TSP at 30 signalized intersections along Route 5 in late 2018 and completed the installation and testing processes in March 2019. By late March 2019, all 30 intersections had TSP activated, and buses enabled to send out requests for priority. The 30 signalized intersections were located along two sections of Route 5 within the central-outer city transition area. The North Section was from North 7th Street and Olsen Memorial Highway to Fremont Avenue North and 42nd Avenue North, with 18 TSP intersections. The South Section was from Chicago Avenue and East Franklin Avenue to Chicago Avenue and East 54th Street, with 12 TSP intersections.

TSP equipment and system for Route 5 were provided and installed by EMTRAC. The TSP system uses buses' GPS location and schedule adherence status to determine the activation of signal priority, as illustrated in Figure 1. A Virtual "check-in" detection zone was set up to cover a range from about a block upstream to the stop line of each signalized intersection. A shorter "check-out" detection zone was set up on the immediate downstream of the stop line. A bus entering a "check-in" detection zone sends out a request for priority. An EMTRAC detector, connected with a signal controller, determines whether to grant priority, and how much green time extension or red time truncation to apply. When a bus having received priority clears the intersection, the "check-out" detection is activated, sending a message to the signal controller. The signal controller ends the priority and enters a recovery process in succeeding cycles.

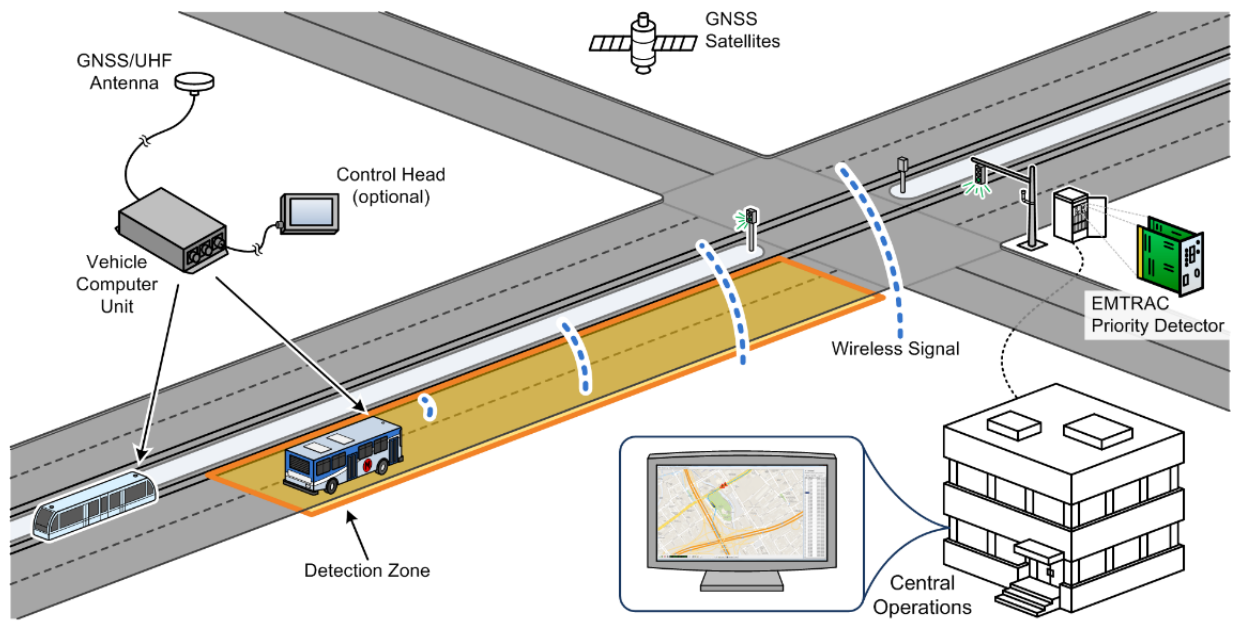

**Figure 1  EMTRAC GPS-based TSP (EMTRAC 2017)**

TSP implementations on Route 5 did not involve stop-relocation or other changes in the bus routes' facilities. The only geometric changes took place on the one-way thoroughfare of Fremont Avenue North and Emerson Avenue North between Plymouth Avenue North and North 33rd Avenue. The changes include bike lane conversion and sidewalk bulb-out (i.e., curb extension), which were implemented around the same time with TSP installation. All geometric changes were considered in this analysis, with those factors' effects controlled and not included in the estimation of TSP's effect.

**DATA**

Data used in this study were provided by Metro Transit. Except for the geometric change information, all data were collected for Route 5 from the months of April-June in 2018 and 2019, to ensure similar climate and travel patterns for the before and after periods.

**Bus route, schedule, TSP activation status, and geometric features.** Bus route map and schedules for Route 5's weekday, Saturday, and Sunday services were provided by Metro Transit. The bus schedules were adjusted several times during both the before and after periods. The schedules were used to validate bus trip and timestamp information in the AVL data set. TSP activation status was provided. All 30 intersections evaluated in this study had TSP installed and activated by March 2019. Information about geometric design features including one-way street design, nearside bus stop, number of lanes, roadside parking, bike lanes, bulb-outs, and curb cuts were collected through historical Google Maps Street Views and relevant Metropolitan Council street improvement project documents. Those features were all coded and organized by bus operating direction (i.e., northbound and southbound) in an intersection summary data set. All geometric changes along Emerson Avenue North and Fremont Avenue North were coded based on the Emerson & Fremont Pedestrian & Bicycle Enhancements Project CAD drawings from Alliant Engineering and City of Minneapolis Public Works, who designed and implemented the geometric changes.

**Automatic vehicle location (AVL).** AVL data for the April-June months of 2018 and 2019 were provided, with records of all Route 5 buses in operation during the periods. Data elements included date, timestamp, bus ID, block ID, trip ID, vehicle direction, latitude and longitude of coordinates, and schedule adherence (in minutes). Route 5 bus AVL data were collected by an on-board device every 8 seconds. The AVL data offered information about bus movements and activities at intersections and its upstream and downstream road segments.

**TSP requests.** A total of 81,392 TSP requests were sent out by Route 5 buses during April-June 2019. Date, timestamp, intersection ID, bus ID, and direction were included in the TSP request data set.

**METHODOLOGY**

An event-level analysis was carried out to evaluate the relationship between TSP request and a surrogate safety measure of bus speed volatility (BSV). For this study, an event is defined as a bus traversal (i.e., approaching and clearing) of a TSP intersection. For both operating directions of the 30 TSP intersections on Route 5, there were over 4,200 such events taking place every day. Linear regression was used to model the relationship between BSV and TSP request, as well as other impact factors.

**BSV as a surrogate safety measure.** From a probabilistic viewpoint, Davis et al. proposed a conceptual framework showing the relationship between surrogate measures and safety outcomes (Davis et al. 2011). The framework implies that some measurable evasive actions, together with initial conditions (including environmental and human factors), determines the probability of the occurrence of a safety outcome such as a traffic crash or a passenger incident. Therefore, by using measurable variables describing initial conditions, and measurable variables describing road user actions, risks of safety outcomes can be estimated. Measures such as time to collision (TTC), post-encroachment time (PET), and deceleration rate are common safety surrogates used for safety risk estimation (Johnsson et al. 2018). Depending on the interested road user groups, safety surrogates have been developed for specific groups such as pedestrians. Combinations of surrogate indicators have also been developed to capture both the probability of crash occurrence and the probabilities of different crash severity levels.

Previous studies have also used vehicle trajectory, speed profile, or driving volatility as surrogate measures of safety risks (Arvin et al. 2019; Boonsiripant 2009; Kachroo and Sharma 2018; Kamrani et al. 2018; Moreno and García 2013; Peesapati et al. 2011; Wali et al. 2018). Changes in speed can reflect the operating status of a vehicle. When an emergency situation occurs or a crash happens, vehicle speed and trajectory are expected to change dramatically. BSV is a quantitative measure of bus speed changes, or volatility. Volatility is a concept frequently used in finance, especially in describing changes in stock market prices. There are many different measures of stock volatility, and coefficient of variation is one of them. A coefficient of variation measures the dispersion of data points in a data series around the mean (Hayes 2020). BSV is a coefficient of variation for bus speeds, measuring the dispersion of bus interval speeds (recorded every 8 seconds by AVL system) around the mean speed of a bus through an event of going from the upstream to the downstream of a signalized intersection. A larger BSV means more dramatic changes in bus speed, thus more safety risks. BSV was calculated as:

$$BSV = \frac{s.d.(v)}{abs(mean(v))}$$

where: *BSV* = bus speed volatility;
   *v* = interval speed (mph);
   *s.d.* = standard deviation;
   *abs* = absolute value; and
   *mean* = mean value.

**Data processing.** To formulate a data set suitable for statistical analysis, the raw data sets were cleaned and merged following the procedure shown in Figure 2. The AVL data was filtered using intersection upstream and downstream limits to get event-level data including variables such as calculated BSV, upstream approach speed, and schedule adherence. The upstream and downstream limits of an intersection are determined by the limits of TSP activation zones. About one block upstream and one block downstream from the intersection was bordered by the limits. Since the TSP intersections were mostly spaced more than one block away from each other, there is no overlap between the buffers bordered by the limits. The upstream approach speed was calculated using the first three AVL coordinates on each time-space trajectory of a bus event. Merging with intersection attributes, 30 intersections' event-level data sets were combined. The combined event-level data set was then merged with TSP request data, using intersection ID, operating direction, date, and bus ID as the matching criteria. The merged final event-level data set was ready to use for statistical analysis.

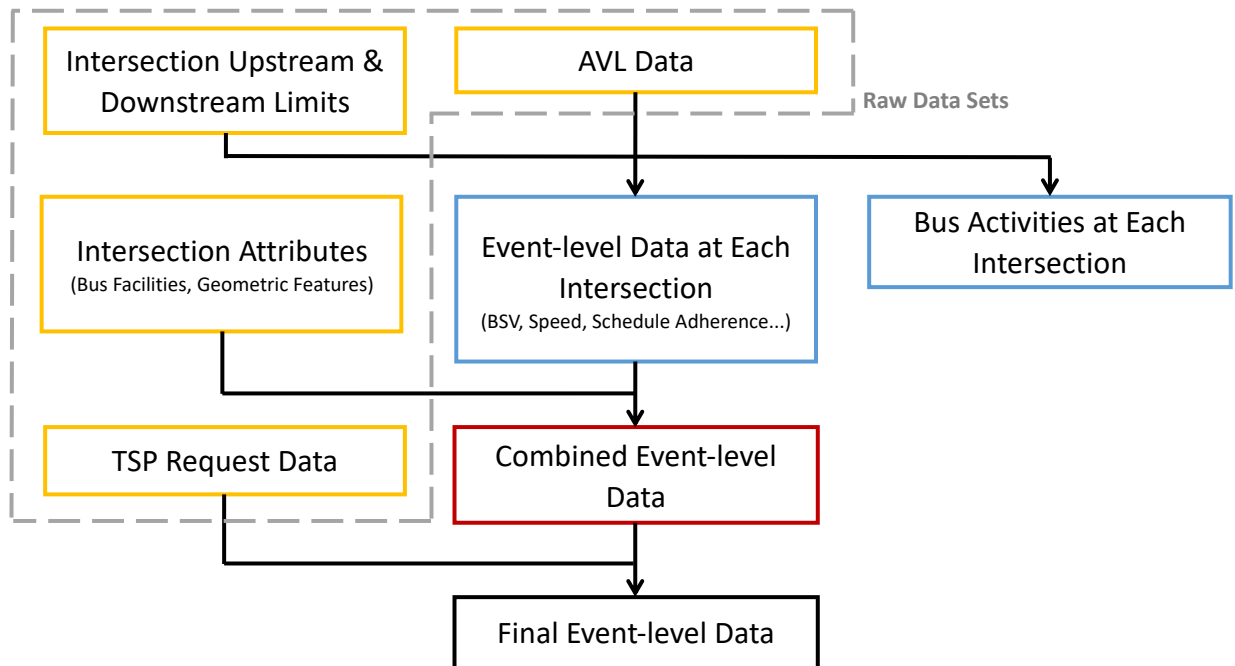

**Figure 2  Data processing procedure**

   To reduce potential bias in modeling due to outliers in the calculated measures such as BSV and approach speed, extremely large values were excluded from these two variables.

The cleaned final data set consists of 23,123 rows of event observations and 23 variables, covering all 30 TSP intersections, as well as the April-June periods of 2018 and 2019.

The descriptive statistics of all variables in the final data set were summarized in Table 1. BSV was the outcome variable of interest, ranging from 0.00 to 2.69, with a mean value of 0.90. Candidate explanatory variables were grouped into three categories, bus operations-related variables, roadway and intersection-related variables, and high-incident intersection indicators. In about 8% of the 23,123 events, TSP was requested. High-crash/incident intersections were those with the highest numbers of bus crashes and passenger incidents during April-June 2018.

Table 1 Descriptive statistics

| Variable | Description | Min | Max | Mean | S.D. |
|---|---|---|---|---|---|
| bsv | Bus speed volatility | 0.00 | 2.69 | 0.90 | 0.57 |
| **Bus Operations** | | | | | |
| after | Record from after period (2019) (yes=1, no=0) | 0.00 | 1.00 | 0.46 | 0.50 |
| tsp_req | Sent TSP request (yes=1, no=0) | 0.00 | 1.00 | 0.08 | 0.27 |
| dir | Direction (southbound=1, northbound=0) | 0.00 | 1.00 | 0.45 | 0.50 |
| spd | Upstream bus approach speed (mph) | 0.00 | 59.89 | 21.70 | 11.78 |
| adh | Bus schedule adherence (min) | -60.00 | 22.00 | -3.98 | 5.01 |
| late | Bus late arrival (if adh ≤ -1; yes=1, no=0) | 0.00 | 1.00 | 0.78 | 0.41 |
| **Roadway and Intersection Geometries** | | | | | |
| oneway | One-way street (yes=1, no=0) | 0.00 | 1.00 | 0.14 | 0.34 |
| nearside | Nearside bus stop (yes=1, no=0) | 0.00 | 1.00 | 0.65 | 0.48 |
| numlanes | Number of lanes on road segment | 1.00 | 2.00 | 1.06 | 0.23 |
| totlane | Number of lanes at intersection approach | 1.00 | 3.00 | 1.85 | 0.47 |
| ltlane | Number of left-turn lanes | 0.00 | 1.00 | 0.50 | 0.50 |
| rtlane | Number of right-turn lanes | 0.00 | 1.00 | 0.29 | 0.45 |
| bulbout | Bulb-out at intersection (yes=1, no=0) | 0.00 | 1.00 | 0.03 | 0.18 |
| sideparking | Road-side parking within a block (yes=1, no=0) | 0.00 | 1.00 | 0.85 | 0.36 |
| rightbike | Right-side bike lane (yes=1, no=0) | 0.00 | 1.00 | 0.44 | 0.50 |
| leftbike | Left-side bike lane (yes=1, no=0) | 0.00 | 1.00 | 0.10 | 0.30 |
| curbcut | Curb cut within a block (yes=1, no=0) | 0.00 | 1.00 | 0.42 | 0.49 |
| **High-Crash/Incident Intersection Indicators** | | | | | |
| int62 | Emerson & Broadway (yes=1, no=0) | 0.00 | 1.00 | 0.01 | 0.10 |
| int67 | 7th & Olson (yes=1, no=0) | 0.00 | 1.00 | 0.00 | 0.05 |
| int73 | Chicago & Lake (yes=1, no=0) | 0.00 | 1.00 | 0.08 | 0.27 |

Note: Number of observations: 23,123.
Min = minimum; Max = maximum; S.D. = standard deviation.

**Linear regression.** A regression analysis was carried out to estimate the effect of TSP request on buses' safety risks, measured by BSV. Since the outcome variable, BSV, is continuous, linear regression was used as the analytical method, modeling the outcome variable as a linear combination of the explanatory variables:

$$y = X\beta + \varepsilon$$

where: $y$ = a vector of observed values for the outcome variable;
$X$ = an n-dimensional column vector of explanatory variables, and n = number of observations;
$\beta$ = a vector of estimated parameters, with $\beta_0$ being an intercept term and $\beta_1, \beta_2, \ldots$ being estimated parameters for explanatory variables; and
$\varepsilon$ = a vector of error terms.

A linear regression model has five assumptions about the generation of observations (Kennedy 2008):
- The linear relationship between outcome variable and the set of explanatory variables;
- The expected value of the error term is zero;
- The error terms all have the same variance and are not correlated;
- The observations on the explanatory variables can be considered fixed in repeated samples;
- The number of observations is greater than the number of explanatory variables and there are no exact linear relationships between any pair of the explanatory variables.

These assumptions were checked through model diagnostics. Before conducting linear regression, the correlations between each pair of the 21 variables (1 outcome variable + 20 explanatory variables) are checked, with no strong correlations (> 0.70) found except for a relatively higher correlation between one-way street and left-side bike lane (0.83). That is because all left-side bike lanes are on the one-way road segments in the sample. Therefore, the left-side bike lane variable was dropped from modeling since the one-way street variable is of more interest.

**RESULTS**

Model results were summarized in Table 2. The final model included 17 explanatory variables. All 23,123 observations in the final data set were used for modeling. The model is statistically significant, with an F statistic (on 17 variables with 23,105 degrees of freedom) of 652.7 ($p < 0.000$, $R^2 = 0.324$). The model results show that BSV is significantly associated with TSP request ($p < 0.01$). With TSP requested, BSV was lower by about 0.04, compared with when TSP was not requested. With TSP requested, a bus moved through an intersection more smoothly than without TSP requested.

Coefficient estimates of other variables also provide insights about other factors affecting the surrogate safety measure. The overall BSV did not change significantly from 2018 to 2019 ($\beta\_after = 0.003$, $p > 0.1$). Southbound buses had a higher BSV than northbound buses ($\beta\_dir = 0.119$, $p < 0.001$). A higher approach speed is associated with a lower BSV ($\beta\_spd = -0.014$, $p < 0.001$). Late arrival is associated with a higher BSV compared with on-time arrival ($\beta\_late = 0.131$, $p < 0.001$). BSV is also significantly related to roadway and intersection geometric features ($p < 0.001$). One-way street design, more lanes, the existence of a turn lane, and road-side parking are associated with a higher BSV. A near-side bus stop, sidewalk bulb-out, right-side bike lane, and curb cuts are associated with a lower BSV. The intersections with higher numbers of bus-involved crashes and passenger incidents are all significantly associated with a higher BSV ($\beta\_int62 = 0.094$, $p < 0.01$; $\beta\_int67 = 0.646$, $p < 0.001$; $\beta\_int73 = 0.550$, $p < 0.001$), compared with other intersections.

Table 2  Linear regression model

| Variable | Description | Estimate | S.E. | t-value | Significance |
|---|---|---|---|---|---|
| (Intercept) | | 1.058 | 0.026 | 40.75 | *** |
| **Bus Operations** | | | | | |
| after | 2019 indicator | 0.003 | 0.007 | 0.41 | |
| tsp_req | TSP request | -0.038 | 0.012 | -3.11 | ** |
| dir | Direction | 0.119 | 0.008 | 15.45 | *** |
| spd | Approach speed | -0.014 | 0.000 | -49.13 | *** |
| late | Late arrival | 0.131 | 0.008 | 16.79 | *** |
| **Roadway and Intersection Geometries** | | | | | |
| oneway | One-way street | 0.056 | 0.013 | 4.26 | *** |
| nearside | Nearside bus stop | -0.288 | 0.010 | -29.22 | *** |
| numlanes | No. of lanes (segment) | 0.145 | 0.016 | 9.01 | *** |
| ltlane | Left-turn lane | 0.048 | 0.010 | 4.68 | *** |
| rtlane | Right-turn lane | 0.068 | 0.014 | 4.90 | *** |
| bulbout | Bulb-out | -0.442 | 0.021 | -21.07 | *** |
| sideparking | Road-side parking | 0.104 | 0.014 | 7.54 | *** |
| rightbike | Right-side bike lane | -0.200 | 0.010 | -19.37 | *** |
| curbcut | Curb cut (driveway) | -0.119 | 0.007 | -16.29 | *** |
| **High-Crash/Incident Intersections** | | | | | |
| int62 | Emerson & Broadway | 0.099 | 0.035 | 2.83 | ** |
| int67 | 7th & Olson | 0.648 | 0.070 | 9.31 | *** |
| int73 | Chicago & Lake | 0.550 | 0.016 | 33.41 | *** |

R-squared: 0.324
Adjusted R-squared: 0.324
F-statistic: F(df: 17 & 23,105) = 652.7***

Note: S.E. = standard error.
Significance level: "***" 0.001; "**" 0.01; "*" 0.05; "." 0.1; " " 1.

Model diagnostics were supported by plots in Figure 3. There were four plots in each diagnostic set, the residuals vs. fitted plot, the normal Q-Q (quantile-quantile) plot, the scale-location plot, and the residual vs. leverage plot. The four plots were used to check the assumptions about linearity, residual normality, residual homoscedasticity, and extreme and influential cases (Kassambara 2018). The residual plot showed no obvious patterns between residuals and fitted values from the model. Thus, the assumption about linear relationship between the outcome variable and explanatory variables was supported. The Q-Q plot showed that normal probabilities of residuals approximately followed the reference line, confirming the residual normality assumption. The scale-location plot showed that the variances of the residual points did not change much when the value of the fitted value changes. Therefore, the scale-location plot supported the residual homoscedasticity assumption, that the error terms all had the same variance and were not correlated. The residual vs. leverage plot showed that some points had larger standardized residuals and some higher leverage values, but with a large sample size those phenomena were expected. No influential points were shown on the plot, with no Cook's distance lines shown. Therefore, the diagnostic for possible outliers and high-leverage points was acceptable.

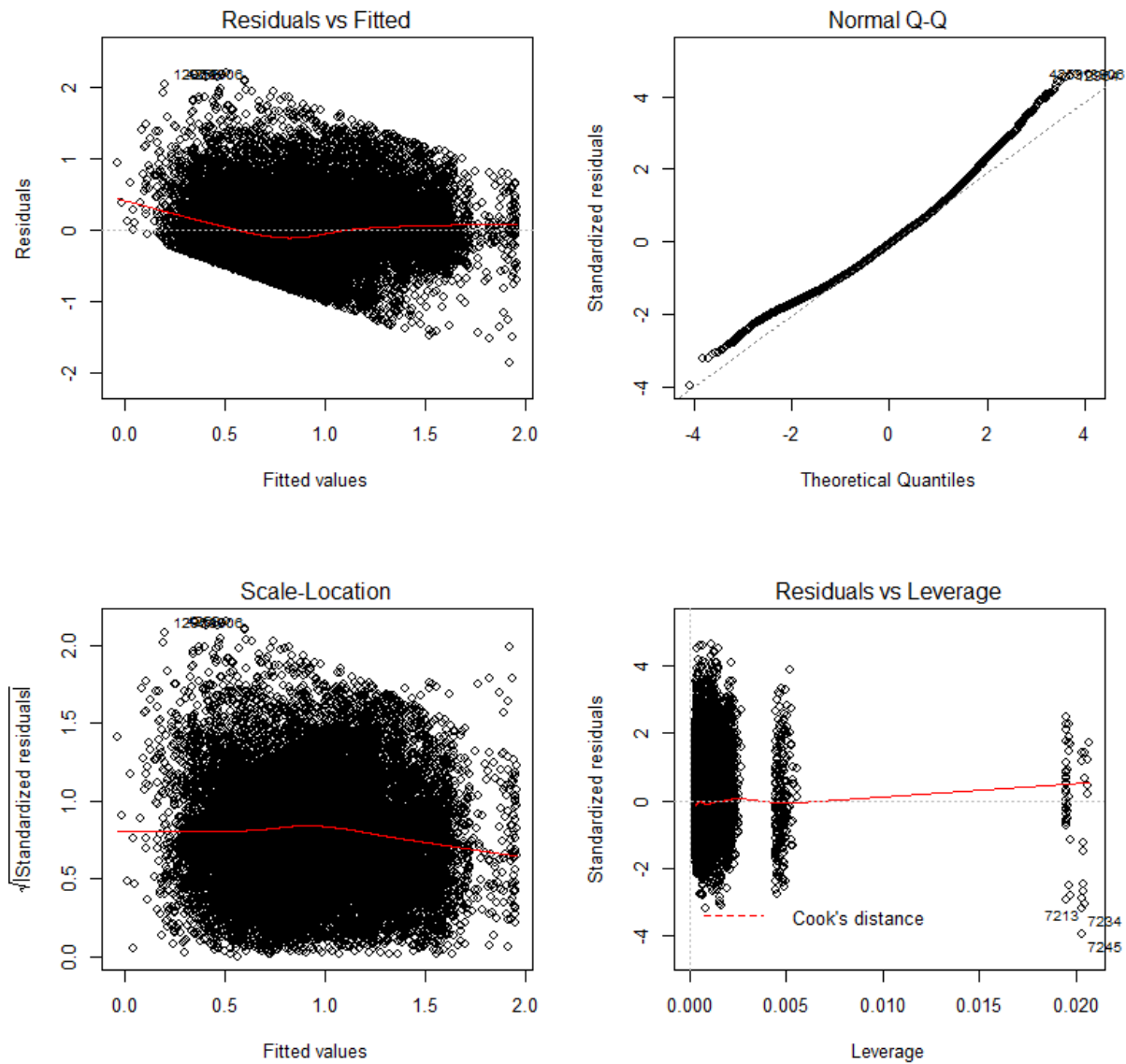

**Figure 3  Linear regression diagnostics**

## CONCLUSIONS

With detailed TSP and bus operational data from Minneapolis-St. Paul Metro Transit, a statistical analysis was completed to evaluate TSP's effects on bus safety risks, measured by BSV. The linear regression analysis results showed that TSP requests were significantly associated with lower BSV through signalized intersections. With TSP requested, buses could traverse TSP-equipped intersections more smoothly without many harsh accelerations, deceleration, or stops. This finding supports the hypothesis that TSP affects safety through its effects on bus movements. This finding once again indicates that a benefit of safety risk reduction could be expected from implementing TSP, as smoother bus movements are closely linked to safer and more comfortable bus trips. A cumulated benefit from safer bus trips is a safer bus corridor. This finding is consistent with those from most previous studies on TSP's safety effects.

In addition to TSP requests, many other factors also showed significant relationships with BSV. Analysis results indicated that bus operational and roadway/intersection geometric factors also affect bus safety significantly and implied that those factors should not be overlooked when evaluating TSP projects' potential safety effects.

There were some limitations to this study that can be addressed in future work. As TSP request records were used to distinguish buses that potentially benefited from TSP, we were not sure about which buses were granted TSP. Signal activity logs were not stored for a period long enough for this study. There existed a probability of buses requested TSP but were not granted TSP by the signal controllers, considering which, the TSP's safety effects might be slightly under-estimated in this study. Although only one measure, BSV, was developed and used in this study, other surrogate safety measures can be explored and used based on future TSP safety evaluation needs.